\documentclass[12pt,a4paper]{article}
\usepackage{a4wide}
\usepackage{amsmath}
\usepackage{amssymb}
\usepackage{epsfig}
\usepackage{subfigure}
\usepackage{exscale}
\usepackage{float}
\usepackage{bbm}
\usepackage[numbers,sort&compress]{natbib}
\usepackage{pst-plot, pstricks,pst-math}
\usepackage{fancybox,amssymb,color}
\usepackage{graphicx}
\usepackage{pstricks, color, graphicx, epsfig, psfrag}
\usepackage{amsfonts,amsmath,amssymb,slashed}
\usepackage{dsfont}
\usepackage{bbm,bm}
\usepackage{fancyhdr, a4wide}
\usepackage[english]{babel}
\usepackage{subfigure}

\makeatletter
\@addtoreset{equation}{section}
\makeatother

\title{Quark Condensate in a Weak Magnetic Field}

\author{Christoph P.\ Hofmann$^a$ \\ \\
\normalsize{$^a$ Facultad de Ciencias, Universidad de Colima} \\
\vspace{0.3cm}
\normalsize {Bernal D\'iaz del Castillo 340, Colima C.P.\ 28045, Mexico} \\}

\begin{document}

\maketitle

\begin{abstract} \normalsize

The low-temperature representation for the quark condensate in a weak magnetic field $H$ is known up to two-loop order. Remarkably, at
one-loop order, the published series for the quark condensate in the chiral limit and $H \ll T^2$ are inconsistent. Using an alternative
representation for the kinematical Bose functions, we derive the series to arbitrary order in $H/T^2$, and also determine which of the
published results is correct.

\end{abstract}

\maketitle

\section{Motivation}
\label{Intro}

The low-energy behavior of quantum chromodynamics in the presence of a magnetic field has been explored by many authors in great detail.
The partition function has been evaluated up to two-loop order within chiral perturbation theory, and low-temperature series for the quark
condensate have been presented. A comprehensive list of references can be found in the nice review by Andersen et al. \citep{ANT16}.

Of particular interest is the low-temperature expansion of the quark condensate in the chiral limit in a {\it weak} magnetic field $H$. The
weak magnetic field limit is implemented by $|qH| \ll T^2$ where $|q|$ stands for the electric charge of the pion. The relevant quantity
is 
\begin{equation}
\frac{\langle {\bar q} q \rangle}{\langle 0 | {\bar q} q | 0 \rangle} \, ,
\end{equation}
where $\langle 0 | {\bar q} q | 0 \rangle$ is the quark condensate at zero temperature and zero magnetic field. Two series are available
in the literature for the above quantity in the chiral limit: up to one-loop order, the author of Refs.~\citep{Aga00,Aga01,AS01,Aga08}
obtains
\begin{eqnarray}
\label{agasian}
\frac{\langle {\bar q} q \rangle}{\langle 0 | {\bar q} q | 0 \rangle} & = & 1 + \frac{{\cal C} \, |qH|}{16 \pi^2 F^2} - \frac{T^2}{8 F^2}
- \frac{7 \sqrt{|qH|} T}{48 \pi F^2} - \frac{|qH|}{16 \pi^2 F^2} \, \log{\frac{|qH|}{T^2}} \, , \qquad |qH| \ll T^2 \, , \nonumber \\
& & {\cal C} = \log 2 - 2 \gamma_E + 2 \log 4 \pi + \frac{1}{3} \, ,
\end{eqnarray}
while the author of Refs.~\citep{And12a,And12b} ends up with
\begin{equation}
\label{andersen}
\frac{\langle {\bar q} q \rangle}{\langle 0 | {\bar q} q | 0 \rangle} = 1 + \frac{|qH| \log 2}{16 \pi^2 F^2} - \frac{T^2}{8 F^2}
+ \frac{5 \sqrt{|qH|} T}{48 \pi F^2} + \dots \, , \qquad |qH| \ll T^2 \, .
\end{equation}
Both series contain the leading term at zero temperature,
\begin{equation}
\frac{|qH| \log 2}{16 \pi^2 F^2} \, ,
\end{equation}
which is linear in the magnetic field and positive, and has been derived in the pioneering paper by Shushpanov and Smilga
\citep{SS97}.\footnote{Note that the constant ${\cal C}$ in Eq.~(\ref{agasian}) involves further terms:
$- 2 \gamma_E + 2 \log 4 \pi + \frac{1}{3}$. As we comment at the end of Section \ref{evaluation}, these terms do not contribute at zero
temperature.} As far as finite-temperature corrections are concerned, we first have a term that does not involve the magnetic field,
\begin{equation}
- \frac{T^2}{8 F^2} \, ,
\end{equation}
derived a long time ago in the original article by Gasser and Leutwyler ~\citep{GL87}. However, in nonzero magnetic field, the two series
disagree with respect to the leading contribution at finite temperature: the coefficients of the $\sqrt{H} T$-term are different both in
magnitude and sign. Finally, according to Refs.~\citep{Aga00,Aga01,AS01,Aga08}, logarithmic terms of the form $H \log(H/T^2)$ also emerge.

In order to make the low-temperature expansion in the weak magnetic field limit ($|qH| \ll T^2$) more transparent, we factorize out
temperature and use the relevant expansion parameter $\epsilon < 1$,
\begin{equation}
\epsilon = \frac{|qH|}{T^2} \, .
\end{equation}
The two published series can then be cast into the general form
\begin{eqnarray}
\frac{\langle {\bar q} q \rangle}{\langle 0 | {\bar q} q | 0 \rangle} & = & 1 + \frac{|qH| \log 2}{16 \pi^2 F^2}
+ \Big\{ q_1 \sqrt{\epsilon} + q_2 \, \epsilon \log \epsilon + q_3 \, \epsilon + q_4 \, \epsilon^2 + q_5 \, \epsilon^3
+ {\cal O}(\epsilon^4) \Big\} \, T^2 \nonumber \\
& & - \frac{1}{8 F^2} \, T^2 + {\cal O}(T^4) \, .
\end{eqnarray}
Let us consider the quantity
\begin{equation}
Q(\epsilon) = \frac{1}{\sqrt{\epsilon} \, T^2} \, \Bigg( \frac{\langle {\bar q} q \rangle}{\langle 0 | {\bar q} q | 0 \rangle} - 1
- \frac{|qH| \log 2}{16 \pi^2 F^2} + \frac{1}{8 F^2} \, T^2 - \, {\cal O}(T^4)\Bigg) \, .
\end{equation}
In the limit $\epsilon \to 0$, we have
\begin{equation}
\lim_{\epsilon \to 0} Q(\epsilon) = q_1 \, .
\end{equation}
Irrespective of whether or not a logarithmic contribution is present, $Q(\epsilon)$ should converge to the leading coefficient $q_1$. The
authors of Refs.~\citep{Aga00,Aga01,AS01,Aga08} and Refs.~\citep{And12a,And12b} end up with different values for $q_1$. The motivation for
the present study is to decide which of the two published results is correct, and to go to higher orders in the weak magnetic field
expansion. Our calculation is based on chiral perturbation theory, much like Refs.~\citep{Aga00,Aga01,AS01,Aga08,And12a,And12b}, but
relies on an alternative representation for the kinematical Bose functions that appear at one-loop order -- our approach then allows for a
systematic and very transparent expansion in the limit $|qH| \ll T^2$.

As it turns out, our leading coefficient $q_1$ is yet different from the two published results, and higher-order terms in our series
disagree with the Agasian series \citep{Aga00,Aga01,AS01,Aga08}. We have checked that our series perfectly coincides with the exact result
that we have evaluated numerically. We stress that the criticism is not directed towards the one-loop evaluation of the partition function
-- rather, our intention is to point out that, in the low-temperature expansion of the quark condensate at one-loop order, errors exist
concerning the weak magnetic field expansion $|qH| \ll T^2$ in the chiral limit. More important, the correct series is derived in the
present study for the first time.

\section{Quark Condensate in Weak Magnetic Fields}
\label{evaluation}

The essentials of chiral perturbation theory have been outlined in many excellent reviews where the interested reader is referred to (see,
e.g., Refs.~\citep{Leu95,Sch03}). Here we merely provide a brief sketch of the method and the one-loop evaluation.

The QCD Lagrangian for two flavors reads
\begin{equation}
{\cal L}_{QCD} = - \frac{1}{2g^2} \mbox{tr}_cG_{\mu\nu} G^{\mu\nu} + \bar{q} i \gamma^\mu D_\mu q - \bar{q}\, m\,q \, , \qquad (q = u, d) \, .
\end{equation}
In the present study we focus on the isospin limit $m_u = m_d$. The quark condensate,
\begin{equation}
\langle 0 | \, {\bar q} q \, | 0 \rangle \, ,
\end{equation}
is the order parameter associated with the spontaneously broken chiral symmetry $SU(2) \times SU(2) \to SU(2)$. The corresponding
Goldstone bosons are the three pions.

In the effective field theory, the pion fields $\pi^i \, (i=1,2,3)$ are contained in the SU(2) matrix $U=\exp(i \tau^i \pi^i/F)$, where
$\tau^i$ are the Pauli matrices and $F$ is the (tree-level) pion decay constant. The leading term in the effective Lagrangian is of
momentum order $p^2$ and reads
\begin{equation}
\label{L2}
{\cal L}^2_{eff} = \mbox{$ \frac{1}{4}$} F^2 Tr \Big[ {(D_{\mu} U)}^\dagger (D_{\mu} U) - M^2 (U + U^\dagger) \Big] \, ,
\end{equation}
where $M$ is the (tree-level) pion mass. It should be pointed out that the magnetic field $H$ is taken into account by the covariant
derivative
\begin{equation}
D_{\mu} U = \partial_\mu U + i [Q,U] A^{EM}_\mu \, .
\end{equation}
$Q$ stands for the charge matrix of the quarks, $Q=diag(2/3,-1/3)e$, and the gauge field $A^{EM}_\mu=(0,0,-Hx,0)$ incorporates the constant
magnetic field in Landau gauge \citep{ANT16}. The next-to-leading piece in the effective Lagrangian -- ${\cal L}^4_{eff}$ -- is of momentum
order $p^4$, and involves various next-to-leading order effective constants $l_i$ and $h_i$ that require renormalization (see Appendix
\ref{appendix}). The explicit form of ${\cal L}^4_{eff}$ can be found, e.g., in Refs.~\citep{GL84,Sch03}.

\begin{figure}
\begin{center}
\includegraphics[width=9cm]{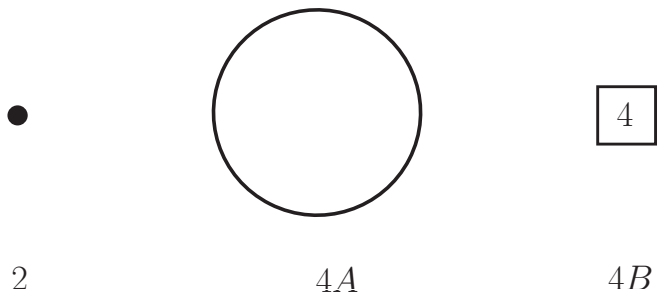}
\end{center}
\caption{QCD partition function diagrams contributing up to one-loop order in the low-temperature expansion. The filled circle refers to
${\cal L}^2_{eff}$, while the number $4$ in the box corresponds to ${\cal L}^4_{eff}$.}
\label{figure1}
\end{figure}

Chiral perturbation theory refers to low temperatures, small quark masses and weak magnetic fields. We first consider the free energy
density from where the quark condensate can be derived. The corresponding Feynman diagrams, up to one-loop order $p^4$, are depicted in
Fig.~\ref{figure1}. Their evaluation leads to the following low-temperature representation for the free energy density,\footnote{The
present evaluation parallels the evaluation of the partition function in zero magnetic field described in much detail in
Refs.~\citep{Hof16a,Hof17}.}
\begin{equation}
\label{freeED}
z = z_0(M,0,H) - \mbox{$ \frac{3}{2}$} g_0(M,T,0) - {\tilde g}_0(M,T,H) + {\cal O} (p^6) \, .
\end{equation}
The quantity $z_0$ is the free energy density at zero temperature, while the two other contributions are finite-temperature corrections:
the first one refers to zero magnetic field, the second one incorporates the magnetic field.

As we show in Appendix \ref{repqq}, the quark condensate can then be obtained from the free energy density. Up to one-loop order we get
\begin{eqnarray}
\label{finalQuarkCondensate}
\frac{\langle {\bar q} q \rangle}{\langle 0 | {\bar q} q | 0 \rangle} & = & 1 - \frac{|qH|}{16 \pi^2 F^2} {\int}_{\!\!\! 0}^{\infty} dt t^{-1}
\Big( \frac{1}{\sinh(t)} - \frac{1}{t} \Big) \nonumber \\
& & - \frac{3 g_1(0,T,0)}{2 F^2} - \frac{{\tilde g_1}(0,T,H)}{F^2} + {\cal O}(p^4) \, ,
\end{eqnarray}
where $\langle 0 | {\bar q} q | 0 \rangle$ is the quark condensate at zero temperature and zero magnetic field. The first line in
Eq.~(\ref{finalQuarkCondensate}) refers to zero temperature, while the kinematical functions $g_1(0,T,0)$ and ${\tilde g_1}(0,T,H)$
describe the behavior of the system at finite temperature. We now analyze in detail the structure of the above terms.

The basic object in the evaluation of the partition function -- or, equivalently, free energy density -- is the thermal propagator $G(x)$
for the pions in the background of a magnetic field. It can be constructed from the zero-temperature propagator $\Delta(x)$ in Euclidean
space by
\begin{equation}
\label{ThermalPropagator}
G(x) = \sum_{n = - \infty}^{\infty} \Delta({\vec x}, x_4 + n \beta) \, , \qquad \beta = \frac{1}{T} \, .
\end{equation}
The propagator $\Delta^0(x)$ referring to the neutral pion is not affected by the magnetic field and takes the simple form
\begin{equation}
\label{regprop}
\Delta^0(x) = (2 \pi)^{-d} \int {\mbox{d}}^d p e^{ipx} (M^2 + p^2)^{-1}
= {\int}_{\!\!\! 0}^{\infty} \mbox{d} \rho (4 \pi \rho)^{-d/2} e^{- \rho M^2 - x^2/{4 \rho}} \, .
\end{equation}
As for the two charged pions, it is convenient to start with the representation for the zero-temperature propagator in Minkowski space
given in Refs.~\citep{ANT16,Sho13},
\begin{eqnarray}
\label{PropagatorJens}
\Delta^{\pm}(x) & = & \exp[i s_{\perp} \Phi(x_{\perp})] \int \frac{d^4 p}{{(2 \pi)}^4} \, e^{-ipx} \Delta^{\pm}(p_{\parallel}, p_{\perp}) \, ,
\nonumber \\
& & \hspace{-1.6cm} \Delta^{\pm}(p_{\parallel}, p_{\perp}) = i {\int}_{\!\!\! 0}^{\infty} \frac{ds}{\cos(|qH|s)} \,
\exp \Bigg(is(p^2_{\parallel} - M^2) - ip^2_{\perp} \frac{\tan(|qH|s)}{|qH|}\Bigg) \, ,
\end{eqnarray}
where
\begin{equation}
\Phi(x_{\perp}) = \frac{|qH|}{2} \, x^1 x^2
\end{equation}
is the so-called Schwinger phase, and the other quantities are
\begin{equation}
p^2_{\parallel} = p^2_0 - p^2_3 \, , \qquad p^2_{\perp} = p^2_1 + p^2_2 \, , \qquad s_{\perp} = sign(qH) \, .
\end{equation}
The point is that the summation over the Landau levels -- associated with the magnetic field -- has already been performed in
$\Delta^{\pm}$. In the thermal propagator there is then only one sum left: the one induced by finite temperature. This simplifies the
calculation considerably. After integration over the momenta, and going from Minkowski to Euclidean space, we obtain
\begin{equation}
\Delta^{\pm}(x) = \frac{|qH|}{{(4 \pi)}^{\frac{d}{2}}} \, e^{-s_{\perp} |qH| x_1 x_2/2} \, {\int}_{\!\!\! 0}^{\infty} ds
\frac{e^{-s M^2}}{s \sinh(|qH| s)} \, \exp\Bigg(-\frac{x^2_4 + x^2_3}{4s} - \frac{|qH| (x^2_1 + x^2_2)}{4 \tanh(|qH|s)}\Bigg) \, ,
\end{equation}
from where the thermal propagator for the charged pions can be constructed via Eq.(\ref{ThermalPropagator}).

Up to one-loop order, the thermal propagator $G(x)$ only has to be evaluated at the origin $x$=0, where it can be decomposed into the
$T$=0 contribution and a second piece that refers to finite temperature,
\begin{equation}
\label{decomposition}
G(0) = \Delta(0) + g_1(M,T,H) \, .
\end{equation}
The latter belongs to the class of kinematical Bose functions $g_r(M,T,H)$ defined by
\begin{equation}
\label{BoseMTH}
g_r(M,T,H) = \frac{T^{d-2r-2}}{{(4 \pi)}^{r+1}} \, |qH| {\int}_{\!\!\! 0}^{\infty} dt \frac{t^{r-\frac{d}{2}}}{\sinh ( |qH| t/ 4 \pi T^2 )} \,
\exp\Big( -\frac{M^2}{4 \pi T^2} t \Big) \Bigg[ S \Big( \frac{1}{t} \Big) -1 \Bigg] \, .
\end{equation}
Here $S(z)$ is the Jacobi theta function
\begin{equation}
\label{Jacobi}
S(z) = \sum_{n=-\infty}^{\infty} \exp(- \pi n^2 z) \, .
\end{equation}
The kinematical Bose functions $g_r(M,T,H)$ describe the thermodynamic properties of the pions in presence of magnetic fields. For the
effective theory to be consistent, the quantities $T, M$ and $H$ must be small with respect to the underlying QCD scale $\Lambda \approx
1 \, GeV$. In this study, we are particularly interested in the chiral limit $M \to 0$ and the weak magnetic field limit $|qH| \ll T^2$.

We proceed with the evaluation of the functions $g_r(M,T,H)$. Since the Taylor expansion of the inverse hyperbolic sine starts with
\begin{equation}
\frac{1}{\sinh(t)} = \frac{1}{t} + {\cal O}(t) \, ,
\end{equation}
we perform the following subtraction in the integrand,\footnote{Note that we just subtract and re-add a term.}
\begin{eqnarray}
\label{subtraction}
g_r(M,T,H) & = & \frac{T^{d-2r-2}}{{(4 \pi)}^{r+1}} \, |qH| {\int}_{\!\!\! 0}^{\infty} dt t^{r-\frac{d}{2}} \,
\Bigg( \frac{1}{\sinh(|qH| t /4 \pi T^2)} - \frac{4 \pi T^2}{|qH| t} \Bigg) \nonumber \\
& & \times \, \exp\Big( -\frac{M^2}{4 \pi T^2} t \Big) \Bigg[ S\Big( \frac{1}{t} \Big) -1 \Bigg] \nonumber \\
& & + \frac{T^{d-2r}}{{(4 \pi)}^r} \, {\int}_{\!\!\! 0}^{\infty} dt t^{r-\frac{d}{2}-1} \, \exp\Big( -\frac{M^2}{4 \pi T^2} t \Big)
\Bigg[ S\Big( \frac{1}{t} \Big) -1 \Bigg] \, .
\end{eqnarray}
The second term describes pions in zero magnetic field, and has been evaluated before in Ref.~\citep{GL89},
\begin{equation}
\label{FreeFunctions}
g_r(M, T, 0) = 2 {\int}_{\!\!\! 0}^{\infty} \frac{\mbox{d} \rho}{(4 \pi \rho)^{\frac{d}{2}}} \, {\rho}^{r-1} \, \exp(- \rho M^2)
\sum_{n=1}^{\infty} \exp(- n^2/{4 \rho T^2}) \, .
\end{equation}
We thus consider the first term that depends on the magnetic field,
\begin{eqnarray}
{\tilde g_r}(M, T, H) & = & \frac{|qH|^{\frac{d}{2} -r}}{{(4 \pi)}^{\frac{d}{2}}} \, {\int}_{\!\!\! 0}^{\infty} dt \, t^{r-\frac{d}{2}}
\Big( \frac{1}{\sinh(t)} - \frac{1}{t} \Big) \, \exp\Big( -\frac{M^2}{|qH|} t \Big) \nonumber \\
& & \times \, \Bigg[ S\Big( \frac{|qH|}{4 \pi T^2 t} \Big) -1 \Bigg] \, .
\end{eqnarray}
Since the analysis of ${\tilde g_r}(M, T, H)$ is rather technical, we relegate it to an appendix. In the same appendix \ref{appendix} we
also discuss the structure of the $T$=0 contribution in the free energy density $z_0(M,0,H)$. Here we just provide the final
representation for the quark condensate in the chiral limit and $|qH| \ll T^2$. The latter limit is implemented by expanding the various
quantities in Eq.~(\ref{finalQuarkCondensate}) in the parameter $\epsilon$, 
\begin{equation}
\epsilon = \frac{|qH|}{T^2} \, .
\end{equation}
Up to one-loop order, the low-temperature expansion of the quark condensate in the chiral limit and $|qH| \ll T^2$ then takes the form
\begin{eqnarray}
\label{condensateMySeries}
\frac{\langle {\bar q} q \rangle}{\langle 0 | {\bar q} q | 0 \rangle} & = & 1 + \frac{|qH| \log 2}{16 \pi^2 F^2}
+ \Bigg\{ \frac{|I_{\frac{1}{2}}|}{8 \pi^{3/2} F^2} \sqrt{\epsilon} - \frac{\log 2}{16 \pi^2 F^2} \, \epsilon - \frac{a_1}{F^2} \,
\epsilon^2 - \frac{a_2}{F^2} \, \epsilon^4 + {\cal O}(\epsilon^6) \Bigg\} \, T^2 \, , \nonumber \\
& & - \frac{1}{8 F^2} \, T^2 + {\cal O}(T^4) \, ,
\end{eqnarray}
where
\begin{eqnarray}
I_{\frac{1}{2}} & = & {\int}_{\!\!\! 0}^{\infty} \, dt t^{-1/2} \Big( \frac{1}{\sinh(t)} - \frac{1}{t} \Big) \approx -1.516256 \, , \nonumber \\
a_1 & = & - \frac{\zeta(3)}{384 \pi^4} \, , \qquad a_2 = \frac{7 \zeta(7)}{98 304 \pi^8} \, .
\end{eqnarray}
The analytical representation for the coefficients $a_p$ can be found in the appendix, along with the numerical values for the first few
coefficients $a_1, \dots, a_5$ in Table \ref{table3}. The series at finite temperature in nonzero magnetic field is thus dominated by the
square-root term $\propto \sqrt{\epsilon}$, followed by a term linear in $\epsilon$. The remaining corrections involve even powers of
$\epsilon$.

The temperature-independent contribution in the quark condensate that involves the magnetic field,
\begin{equation}
\frac{|qH| \log 2}{16 \pi^2 F^2} \, ,
\end{equation}
is the Shushpanov-Smilga term derived a long time ago \citep{SS97}, and later confirmed in
Refs.~\citep{Aga00,Aga01,AS01,Aga08,And12a,And12b}, among others. However, comparing our leading temperature-dependent contribution,
\begin{equation}
\label{leadHof}
\frac{\sqrt{|qH|} T}{8 \pi^{3/2} F^2} \, |I_{\frac{1}{2}}| \, ,
\end{equation}
with the respective leading terms in the two published series, Eq.~(\ref{agasian}),
\begin{equation}
- \frac{7 \sqrt{|qH|} T}{48 \pi F^2} \, ,
\end{equation}
and Eq.~(\ref{andersen}),
\begin{equation}
\label{ander}
\frac{5 \sqrt{|qH|} T}{48 \pi F^2} \, ,
\end{equation}
we observe disagreement with either result. Still, it is interesting to note that the leading term obtained by Andersen,
Eq.~(\ref{ander}), numerically almost coincides with ours,
\begin{equation}
\frac{5}{48 \pi} \approx 0.0331573 \, , \qquad \frac {1}{8 \pi^{3/2}} \, |I_{\frac{1}{2}}| \approx 0.0340375 \, ,
\end{equation}
in particular, it is also positive. As far as higher-order contributions are concerned, we cannot confirm the emergence of logarithmic
terms of the form $H \log(H/T^2)$ as suggested in Refs.~\citep{Aga00,Aga01,AS01,Aga08}.

To underline the correctness of our series, we perform some simple numerical tests. First of all, we establish the connection between
our kinematical functions and those in the literature. The representation for the kinematical functions used by the authors of
Refs.~\citep{Aga00,Aga01,AS01,Aga08,And12a,And12b} is the same as the one used in Ref.~\citep{DFR11} where numerical data is available.
The relevant Bose function for the quark condensate in the chiral limit reads
\begin{equation}
R(0,T,H) = \frac{\epsilon \, T^2}{2 \pi} \sum_{k=0}^{\infty} \, {\int}_{\!\!\! -\infty}^{\infty} \frac{dz}{2 \pi} \,
\frac{1}{\sqrt{z^2 + (2k+1) \epsilon }} \, \frac{1}{\exp\Big[ \sqrt{z^2 +(2k+1) \epsilon} \Big] - 1} \, .
\end{equation}
Using Table I of Ref.~\citep{DFR11}, we have verified that the connection between the kinematical function $R(0,T,H)$ and our
representation ${\tilde g_1}(0,T,H)$,
\begin{equation}
{\tilde g_1}(0,T,H) = \frac{\epsilon \, T^2}{16 \pi^2} \, {\int}_{\!\!\! 0}^{\infty} dt t^{-1} \,
\Bigg( \frac{1}{\sinh(\epsilon t /4 \pi)} - \frac{4 \pi}{\epsilon t} \Bigg) \Bigg[ S\Big( \frac{1}{t} \Big) -1 \Bigg] \, ,
\end{equation}
is given by\footnote{Note that the ratio $|qH|/T^2 = \epsilon$, both in $R(0,T,H)$ and ${\tilde g_1}(0,T,H)$, is arbitrary -- we are not
necessarily referring to the weak magnetic field limit $|qH| \ll T^2$.}
\begin{equation}
\label{connection}
\frac{R(0,T,H)}{T^2} - \frac{1}{12} = \frac{{\tilde g_1}(0,T,H)}{T^2} \, .
\end{equation}
The point is that the function $R(0,T,H)$ contains a temperature-dependent contribution that does not involve the magnetic field: this
term has to be subtracted in order to compare with our representation ${\tilde g_1}(0,T,H)$ that describes the purely $H \neq 0$-part by
definition, Eq.~(\ref{subtraction}). Having established equivalence between previous analyses and ours through Eq.~(\ref{connection}), any
discrepancies in the weak magnetic field limit $|qH| \ll T^2$ can be traced back to the expansion of the kinematical functions in the
parameter $\epsilon = |qH|/T^2$.

\begin{table}[ht!]
\centering
\begin{tabular}{|c||c|c|c|c|}
\hline
$\epsilon$ & $-{\tilde g_1}/T^2$ &  ${\cal O}(\sqrt{\epsilon})$ &  ${\cal O}(\epsilon)$ &  ${\cal O}(\epsilon^2)$ \\
\hline
\hline
0.1    &  0.0103249857050    &  0.01076360492    &  0.01032466434    &  0.0103249857058    \\
\hline
0.05   &  0.00739162808212   &  0.007611018032   &  0.007391547742   &  0.00739162808217   \\
\hline
0.01   &  0.00335985989497   &  0.003403750739   &  0.003359856681   &  0.00335985989497   \\
\hline
0.005  &  0.00238486900367   &  0.002406815229   &  0.002384868200   &  0.00238486900367   \\
\hline
0.001  &  0.00107197111872   &  0.001076360492   &  0.001071971087   &  0.00107197111872   \\
\hline 
0.0005 &  0.000758907108297  &  0.0007611018032  &  0.0007589071003  &  0.000758907108297  \\
\hline
0.0001 &  0.000339936133675  &  0.0003403750739  &  0.0003399361334  &  0.000339936133675  \\
\hline
\end{tabular}
\caption{\it Leading terms in our series (\ref{condensateMySeries}) for the finite-temperature quark condensate in the limit
$|qH| \ll T^2$. The $\sqrt{\epsilon}$-term provides a very good approximation for the exact result, and the series converges rapidly.}
\label{table1}
\end{table}

The first numerical test consists in comparing our series with the exact result. More precisely, we consider successive approximations in
the brace\footnote{Note that each term in the brace has to be multiplied by a factor of $F^2$ in order to make the comparison with
$-{\tilde g_1}/T^2$ that is dimensionless.}
\begin{equation}
\Bigg\{ \frac{|I_{\frac{1}{2}}|}{8 \pi^{3/2} F^2} \sqrt{\epsilon} - \frac{\log 2}{16 \pi^2 F^2} \, \epsilon - \frac{a_1}{F^2} \,
\epsilon^2 + {\cal O}(\epsilon^4) \Bigg\}
\end{equation}
of the expansion (\ref{condensateMySeries}), and compare them with the exact result given by the Bose function
\begin{equation}
- \frac{{\tilde g_1}(0,T,H)}{T^2} \, .
\end{equation}
In Table \ref{table1} we provide numerical data from the series (\ref{condensateMySeries}) by including terms up to order
${\cal O}(\sqrt{\epsilon}), \, {\cal O}(\epsilon)$, and ${\cal O}(\epsilon^2)$, respectively. We notice a clear hierarchy: the
$\sqrt{\epsilon}$-term yields a very good leading approximation, while subsequent terms are heavily suppressed -- our series hence
converges very fast.

\begin{table}[ht!]
\centering
\begin{tabular}{|c||c|c|c|}
\hline
$\epsilon$ & $-{\tilde g_1}/T^2$ & ${\cal O}(\epsilon)$ [Ref.~A] & ${\cal O}(\epsilon)$ [Eq.~(\ref{condensateMySeries})] \\
\hline
\hline
0.1    &  0.0103249857050    &  -0.01053561121    &  0.01032466434    \\
\hline
0.05   &  0.00739162808212   &  -0.008088528975   &  0.007391547742   \\
\hline
0.01   &  0.00335985989497   &  -0.004081832037   &  0.003359856681   \\
\hline
0.005  &  0.00238486900367   &  -0.002980362639   &  0.002384868200   \\
\hline
0.001  &  0.00107197111872   &  -0.001397335349   &  0.001071971087   \\
\hline
0.0005 &  0.000758907108297  &  -0.001000492338   &  0.0007589071003  \\
\hline
0.0001 &  0.000339936133675  &  -0.0004556837879  &  0.0003399361334  \\
\hline
\end{tabular}
\caption{\it Exact result and two expansions for the quark condensate in the weak magnetic field limit at finite temperature. The series
of Refs.~\citep{Aga00,Aga01,AS01,Aga08} (Ref. A) versus our series (\ref{condensateMySeries}). Both series up to linear order in
$\epsilon$.}
\label{table2}
\end{table}

In a second test we compare our series with the one-loop results in the literature. While Andersen in Refs.~\citep{And12a,And12b} provides
the leading term in the weak magnetic field expansion (which numerically is very close to our result), Agasian in Refs.~\citep{Aga00,Aga01,
AS01,Aga08} furthermore derives higher-order corrections for the limit $|qH| \ll T^2$. In Table \ref{table2} we list the numerical values
obtained from the Agasian series, Eq.(54) of Ref.~\citep{Aga08},\footnote{See footnote 5.}
\begin{equation}
\Bigg\{ -\frac{7}{48 \pi F^2} \sqrt{\epsilon} - \frac{1}{16 \pi^2 F^2} \, \epsilon \log \epsilon
- \frac{2(\gamma_E - \log 4 \pi - \mbox{$\frac{1}{6}$})}{16 \pi^2 F^2} \, \epsilon \Bigg\} \, .
\end{equation}
Since the series includes terms up to ${\cal O}(\epsilon)$, we also go up to linear order in our series (\ref{condensateMySeries}).
Inspecting Table \ref{table2}, one notices that the Agasian series does not correctly describe the quark condensate in the weak magnetic
field limit.

A final remark concerns the structure of the low-temperature series. At zero temperature, our series (\ref{condensateMySeries}) reduces to
the Shushpanov-Smilga term as it should. On the other hand, as one approaches zero temperature, the Agasian series Eq.~(\ref{agasian})
formally reduces to
\begin{equation}
\frac{{\cal C} \, |qH|}{16 \pi^2 F^2} \, , \qquad {\cal C} = \log 2 - 2 \gamma_E + 2 \log 4 \pi + \frac{1}{3} \, ,
\end{equation}
which contradicts the original Shushpanov-Smilga result \citep{SS97}. Moreover, the series -- as it stands in
Refs.~\citep{Aga00,Aga01,AS01,Aga08} -- also diverges as zero temperature is approached, because of the logarithmic contribution.

\section{Conclusions}
\label{conclusions}

Expansions for the quark condensate at low temperatures, small pion masses, and weak magnetic fields have been presented up to two-loop
order in the literature. Still, since discrepancies between two published results concern the one-loop level, the present analysis is
perfectly justified.

We emphasize that our approach is based on an alternative representation for the kinematical Bose functions -- different from the
representations used in Refs.~\citep{Aga00,Aga01,AS01,Aga08,And12a,And12b}. Remarkably, we find that the leading term at finite
temperature in the expansion of the quark condensate in a weak magnetic field ($|qH| \ll T^2$), and in the chiral limit, does not coincide
with either of the two published terms. As far as higher-order corrections are concerned, our approach allows for a systematic derivation
of these contributions, that illuminates the structure of the series. 

The low-temperature series is dominated by a square-root term $\sqrt{|qH|/T^2}$ that is positive, much like the (zero-temperature)
Shushpanov-Smilga term. The next term is linear in $|qH|/T^2$ and negative, while subsequent corrections involve even powers of $|qH|/T^2$.
Higher-order terms are heavily suppressed such that our series converges rapidly.

Invoking the exact one-loop expression for the quark condensate -- valid for arbitrary ratio $|qH|/T^2$ -- we have numerically verified
that our expansion correctly describes the quark condensate in weak magnetic fields. We have also observed that the series published in
Refs.~\citep{Aga00,Aga01,AS01,Aga08} fails to approximate the exact result.

\section*{Acknowledgments}
The author thanks J.\ O.\ Andersen for correspondence and R.\ A.\ S\'aenz for helpful comments.

\begin{appendix}

\section{Explicit Calculations}
\label{appendix}

In this appendix we first discuss the free energy density at zero temperature. We then consider the kinematical Bose functions
${\tilde g_r}(M,T,H)$ in the chiral limit and analyze their behavior in weak magnetic fields ($|qH| \ll T^2$). Collecting results, we
provide the representation for the quark condensate in the chiral limit and $|qH| \ll T^2$. Finally, we show how to extract the leading
terms in the expansion of the quark condensate in a straightforward way.

\subsection{Zero Temperature}

The free energy density at zero temperature, up to one-loop order, amounts to
\begin{eqnarray}
\label{vacuumEnergyDensity}
z_0(M,0,H) & = & - F^2 M^2 - (l_3 + h_1) M^4 + 4 h_2 {|qH|}^2 + \mbox{$ \frac{1}{2}$} M^4 \lambda + I_1 + I_2 + {\cal O}(p^6) \, ,
\nonumber \\
& & I_1 = - \frac{{|qH|}^{\frac{d}{2}}}{{(4 \pi)}^{\frac{d}{2}}} {\int}_{\!\!\! 0}^{\infty} dt t^{-\frac{d}{2}-1} \,
\exp\!\Big( -\frac{M^2}{|qH|} t \Big) \, , \nonumber \\
& & I_2 = - \frac{{|qH|}^{\frac{d}{2}}}{{(4 \pi)}^{\frac{d}{2}}} {\int}_{\!\!\! 0}^{\infty} dt t^{-\frac{d}{2}}
\Big( \frac{1}{\sinh(t)} - \frac{1}{t} \Big) \, \exp\!\Big( -\frac{M^2}{|qH|} t \Big) \, .
\end{eqnarray}
The integral $I_1$ can be written as
\begin{equation}
I_1 = M^4 \lambda - \frac{M^4}{64 \pi^2} \, ,
\end{equation}
where $\lambda$
\begin{eqnarray}
\label{lambda}
\lambda & = & \mbox{$ \frac{1}{2}$} \, (4 \pi)^{-\frac{d}{2}} \, \Gamma(1-{\mbox{$ \frac{1}{2}$}}d) M^{d-4} \nonumber \\
& = & \frac{M^{d-4}}{16{\pi}^2} \, \Bigg[ \frac{1}{d-4} - \mbox{$ \frac{1}{2}$} \{ \ln{4{\pi}} + {\Gamma}'(1) + 1 \}
+ {\cal O}(d\!-\!4) \Bigg]
\end{eqnarray}
contains a pole at $d$=4. It should be stressed that factors of $|qH|$ cancel: the integral $I_1$ does not depend on the magnetic field.
The UV-divergence in $I_1$ -- along with the UV-divergence in the term $\mbox{$ \frac{1}{2}$} M^4 \lambda$ of
Eq.~(\ref{vacuumEnergyDensity}) -- can be absorbed into the next-to-leading order effective constants $l_3$ and $h_1$ in the standard
manner, i.e., in chiral perturbation theory where no magnetic field is present (for details see, e.g., Ref.~\citep{GL84}).

The integral $I_2$, that does depend on the magnetic field, also diverges in the limit $d \to 4$. The singularity is proportional to
${|qH|}^2$ and can be absorbed into the next-to-leading order effective constant $h_2$. Explicitly, we subtract the next Taylor term in
the expansion of the inverse hyperbolic sine in $I_2$, such that the integral
\begin{equation}
- \frac{{|qH|}^{\frac{d}{2}}}{{(4 \pi)}^{\frac{d}{2}}} {\int}_{\!\!\! 0}^{\infty} dt t^{-\frac{d}{2}}
\Big( \frac{1}{\sinh(t)} - \frac{1}{t} + \frac{t}{6} \Big) \, \exp\!\Big( -\frac{M^2}{|qH|} t \Big)
\end{equation}
becomes finite if one approaches the physical dimension $d \to 4$. The remainder,
\begin{equation} 
{\hat I}_2 = \frac{{|qH|}^{\frac{d}{2}}}{6{(4 \pi)}^{\frac{d}{2}}} {\int}_{\!\!\! 0}^{\infty} dt t^{-\frac{d}{2}+1} \,
\exp\!\Big( -\frac{M^2}{|qH|} t \Big) \, ,
\end{equation}
can be expressed in terms of $\lambda$ as
\begin{equation} 
{\hat I}_2 = - \frac{{|qH|}^2}{3} \, \lambda - \frac{{|qH|}^2}{96 \pi^2} \, .
\end{equation}
Gathering results, the renormalized free energy density at zero temperature takes the form
\begin{eqnarray}
z_0(M,0,H) & = & - F^2 M^2 + \frac{M^4}{64 \pi^2} \, ({\overline l_3} - 4{\overline h_1} - 1) + \frac{{|qH|}^2}{96 \pi^2} \,
( {\overline h_2} - 1) \\
& & - \frac{{|qH|}^2}{16 \pi^2} {\int}_{\!\!\! 0}^{\infty} dt t^{-2} \Big( \frac{1}{\sinh(t)} - \frac{1}{t} + \frac{t}{6} \Big) \,
\exp\!\Big( -\frac{M^2}{|qH|} t \Big) + {\cal O}(p^6) \, .\nonumber
\end{eqnarray}
Up to the factors $\gamma_3/32 \pi^2$, $\delta_1/32 \pi^2$, and $\delta_2/32 \pi^2$, the constants $\overline l_3$, $\overline h_1$, and
$\overline h_2$ are the running coupling constants at the fixed renormalization scale $\mu = M_{\pi}$, where $M_{\pi} \approx 139.6 MeV$ is
the physical pion mass (for details see, e.g., Ref.~\citep{GL89}).

\subsection{Finite Temperature}
\label{finT}

We now turn to finite temperature where the kinematical Bose functions
\begin{eqnarray}
{\tilde g_r}(M, T, H) & = & \frac{|qH|^{\frac{d}{2} -r}}{{(4 \pi)}^{\frac{d}{2}}} \, {\int}_{\!\!\! 0}^{\infty} dt \, t^{r-\frac{d}{2}}
\Big( \frac{1}{\sinh(t)} - \frac{1}{t} \Big) \, \exp\Big( -\frac{M^2}{|qH|} t \Big) \nonumber \\
& & \times \, \Bigg[ S\Big( \frac{|qH|}{4 \pi T^2 t} \Big) -1 \Bigg]
\end{eqnarray}
become relevant. Our analysis proceeds along the lines of Ref.~\citep{HL90}. From the very start, we refer to the chiral limit that we
implement by $M \to 0$, while keeping $T$ and $|qH|$ fixed. Changing integration variables, and defining $\epsilon=|qH|/T^2$, we first
write 
\begin{equation}
{\tilde g_r}(0, T, H) = \frac{\epsilon}{{(4 \pi)}^{r+1}} T^{d-2r} \, {\int}_{\!\!\! 0}^{\infty} dt \, t^{-\frac{d}{2}+r}
\Big( \frac{1}{\sinh(\epsilon t/4 \pi)} - \frac{4 \pi}{\epsilon t} \Big) \, \Bigg[ S\Big( \frac{1}{t} \Big) - 1 \Bigg] \, .
\end{equation}
The integral is split into two pieces, namely $0 \le t \le 1$ and $1 \le t < \infty$. In the second interval we use the Jacobi identity
\begin{equation}
S(t) = \frac{1}{\sqrt{t}} \, S(1/t) \, ,
\end{equation}
and change the integration variable $t \to 1/t$. This then leads to
\begin{eqnarray}
\label{ABC}
{\tilde g_r}(0, T, H) & = & \frac{\epsilon}{{(4 \pi)}^{r+1}} T^{d-2r} \, {\int}_{\!\!\! 0}^1 dt \, t^{-\frac{d}{2}+r}
\Big( \frac{1}{\sinh(\epsilon t/4 \pi)} - \frac{4 \pi}{\epsilon t} \Big) \, \Bigg[ S\Big( \frac{1}{t} \Big) - 1 \Bigg] \nonumber \\
& & + \frac{\epsilon}{{(4 \pi)}^{r+1}} T^{d-2r} \, \Big\{ I_A + I_B + I_C \Big\} \, ,
\end{eqnarray}
with
\begin{eqnarray}
I_A & = & {\int}_{\!\!\! 0}^1 dt \, t^{\frac{d}{2}-r-\frac{5}{2}} \Big( \frac{1}{\sinh(\epsilon/4 \pi t)} - \frac{4 \pi t}{\epsilon} \Big)
\Bigg[ S\Big( \frac{1}{t} \Big) - 1 \Bigg] \, , \nonumber \\
I_B & = & {\int}_{\!\!\! 0}^1 dt \, t^{\frac{d}{2}-r-\frac{5}{2}} \Big( \frac{1}{\sinh(\epsilon/4 \pi t)} - \frac{4 \pi t}{\epsilon} \Big) \, ,
\nonumber \\
I_C & = & - {\int}_{\!\!\! 0}^1 dt \, t^{\frac{d}{2}-r-2} \Big( \frac{1}{\sinh(\epsilon /4 \pi t)} - \frac{4 \pi t}{\epsilon} \Big) \, .
\end{eqnarray}
The integral $I_B$ we decompose as
\begin{equation}
I_B = {\int}_{\!\!\! 0}^{\infty} dt \, t^{\frac{d}{2}-r-\frac{5}{2}} \Big( \frac{1}{\sinh(\epsilon/4 \pi t)} - \frac{4 \pi t}{\epsilon} \Big)
 - {\int}_{\!\!\! 1}^{\infty} dt \, t^{\frac{d}{2}-r-\frac{5}{2}} \Big( \frac{1}{\sinh(\epsilon/4 \pi t)} - \frac{4 \pi t}{\epsilon} \Big) \, .
\end{equation}
After a few trivial manipulations, we end up with
\begin{eqnarray}
\label{decompositionIB}
I_B & = & I_{B1} + I_{B2} \, , \nonumber \\
I_{B1} & = & \frac{{\epsilon}^{\frac{d}{2}-r-\frac{3}{2}}}{{(4 \pi)}^{\frac{d}{2}-r-\frac{3}{2}}} \, {\int}_{\!\!\! 0}^{\infty} dt \,
t^{-\frac{d}{2}+r+\frac{1}{2}} \Big( \frac{1}{\sinh(t)} - \frac{1}{t} \Big) \, , \nonumber \\
I_{B2} & = & - {\int}_{\!\!\! 0}^1 dt \, t^{-\frac{d}{2}+r+\frac{1}{2}} \Big( \frac{1}{\sinh(\epsilon t/4 \pi)} - \frac{4 \pi}{\epsilon t} \Big) 
\, .
\end{eqnarray}
The integral $I_C$ is processed in an analogous way, with the result
\begin{eqnarray}
\label{decompositionIC}
I_C & = & I_{C1} + I_{C2} \, , \nonumber \\
I_{C1} & = & - \frac{{\epsilon}^{\frac{d}{2}-r-1}}{{(4 \pi)}^{\frac{d}{2}-r-1}} \, {\int}_{\!\!\! 0}^{\infty} dt \,
t^{-\frac{d}{2}+r} \Big( \frac{1}{\sinh(t)} - \frac{1}{t} \Big) \, , \nonumber \\
I_{C2} & = & {\int}_{\!\!\! 0}^1 dt \, t^{-\frac{d}{2}+r} \Big( \frac{1}{\sinh(\epsilon t/4 \pi)} - \frac{4 \pi}{\epsilon t} \Big) \, .
\end{eqnarray}

\subsection{Representation for the Quark Condensate}
\label{repqq}

We now focus on the quark condensate in the chiral limit, 
\begin{equation}
\langle {\bar q} q \rangle = \langle 0 | {\bar q} q | 0 \rangle {\Bigg[ 1 - \frac{1}{F^2} \, 
\frac{\partial}{\partial M^2} \Big( z - z_0(M,0,0) \Big) \Bigg]}_{M^2=0} \, ,
\end{equation}
where
\begin{equation}
z = z_0(M,0,H) - \mbox{$ \frac{3}{2}$} g_0(M,T,0) - {\tilde g}_0(M,T,H) + {\cal O} (p^6)
\end{equation}
is the (total) free energy density and
\begin{equation}
z_0(M,0,0) = - F^2 M^2 + \frac{M^4}{64 \pi^2} \, ({\overline l_3} - 4{\overline h_1} - 1)
\end{equation}
is the $T$=0 contribution that is independent of the magnetic field. Accordingly, the one-loop representation for the quark condensate in
the chiral limit reads
\begin{eqnarray}
\frac{\langle {\bar q} q \rangle}{\langle 0 | {\bar q} q | 0 \rangle} & = & 1 - \frac{|qH|}{16 \pi^2 F^2} {\int}_{\!\!\! 0}^{\infty} dt t^{-1}
\Big( \frac{1}{\sinh(t)} - \frac{1}{t} \Big) \nonumber \\
& & - \frac{3 g_1(0,T,0)}{2 F^2} - \frac{{\tilde g_1}(0,T,H)}{F^2} + {\cal O}(p^4) \, .
\end{eqnarray}
The first line refers to zero temperature, the second line refers to finite temperature.

Our final task is to explore the weak magnetic field limit $|qH| \ll T^2$ that we obtain by expanding in the parameter $\epsilon$
\begin{equation}
\epsilon = \frac{|qH|}{T^2} \, .
\end{equation}
A common factor in the various integrands considered in subsection \ref{finT} is
\begin{equation}
\frac{1}{\sinh(\epsilon t/4 \pi)} - \frac{4 \pi}{\epsilon t} \, ,
\end{equation}
that we expand into
\begin{equation}
\frac{1}{\sinh(\epsilon t/4 \pi)} - \frac{4 \pi}{\epsilon t} = c_1 t \, \epsilon + c_2 t^3 \epsilon^3 + c_3 t^5 \epsilon^5
+ {\cal O}(\epsilon^7) \, .
\end{equation}
The first few Taylor coefficients read
\begin{eqnarray}
& c_1 & = - \frac{1}{24 \pi} \approx -1.33 \times 10^{-2} \, , \nonumber \\
& c_2 & = \frac{7}{23\,040 \, \pi^3} \approx 9.80 \times 10^{-6} \, , \nonumber \\
& c_3 & = - \frac{31}{15\,482\,880 \, \pi^5} \approx -6.54 \times 10^{-9}\, , \nonumber \\
& c_4 & = \frac{127}{9\,909\,043\,200 \, \pi^7}\approx 4.24 \times 10^{-12} \, , \nonumber \\
& c_5 & = -\frac{73}{896\,909\,967\,360 \, \pi^9} \approx -2.73 \times 10^{-15} \, . 
\end{eqnarray}
Introducing the quantities ${\tilde \alpha}_p, {\hat \alpha}_p$, and $\beta_p$ by
\begin{eqnarray}
{\tilde \alpha}_p & = & {\int}_{\!\!\! 0}^1 dt \, c_p \, t^{2p-2} \, \Bigg[ S\Big( \frac{1}{t} \Big) - 1 \Bigg] \, , \nonumber \\
{\hat \alpha}_p & = & {\int}_{\!\!\! 0}^1 dt \, c_p \, t^{-2p-\frac{1}{2}} \, \Bigg[ S\Big( \frac{1}{t} \Big) - 1 \Bigg] \, , \nonumber \\
\beta_p & = & {\int}_{\!\!\! 0}^1 dt \, c_p \, (t^{-1} - t^{-\frac{1}{2}}) \, t^{2p-1} \, ,
\end{eqnarray}
and defining the coefficients $a_p$ as
\begin{equation}
\label{definitionAp}
a_p = \frac{{\tilde \alpha}_p + {\hat \alpha}_p + \beta_p}{16 \pi^2} \, ,
\end{equation}
the low-temperature representation for the quark condensate in the chiral limit and $|qH| \ll T^2$ then takes the form
\begin{eqnarray}
\label{finqcond}
\frac{\langle {\bar q} q \rangle}{\langle 0 | {\bar q} q | 0 \rangle} & = & 1 + \frac{|qH| \log 2}{16 \pi^2 F^2}
+ \Bigg\{ \frac{|I_{\frac{1}{2}}|}{8 \pi^{3/2} F^2} \sqrt{\epsilon} - \frac{\log 2}{16 \pi^2 F^2} \, \epsilon - \frac{a_1}{F^2} \,
\epsilon^2 - \frac{a_2}{F^2} \, \epsilon^4 + {\cal O}(\epsilon^6) \Bigg\} \, T^2 \, , \nonumber \\
& & - \frac{1}{8 F^2} \, T^2 + {\cal O}(T^4) \, .
\end{eqnarray}
The integral $I_{\frac{1}{2}}$ amounts to
\begin{equation}
\label{I12}
I_{\frac{1}{2}} = {\int}_{\!\!\! 0}^{\infty} \, dt t^{-1/2} \Big( \frac{1}{\sinh(t)} - \frac{1}{t} \Big) \approx -1.516256 \, ,
\end{equation}
while numerical values for the first few coefficients $a_p$ in the above expansion are provided in Table \ref{table3}.

\begin{table}[ht!]
\centering
\begin{tabular}{|c|c|}
\hline
$p$  &  $a_p$ \\
\hline
1  &  -3.21361844712 $\times 10^{-5}$  \\
\hline
2  &   7.56726355863 $\times 10^{-9}$  \\
\hline
3  &  -8.00051395855 $\times 10^{-12}$  \\
\hline
4  &   1.87869037118 $\times 10^{-14}$  \\
\hline
5  &  -7.80774216239 $\times 10^{-17}$  \\
\hline
\end{tabular}
\caption{\it Numerical values for the coefficients $a_p$ defined by Eq.~(\ref{definitionAp}).}
\label{table3}
\end{table}

Processing integrals in the same manner as described in the previous subsection, and using the identity
\begin{equation}
\frac{2}{\pi^{\frac{z}{2}}} \, \Gamma\Big(\frac{z}{2}\Big) \zeta(z) = {\int}_{\!\!\! 0}^{\infty} dt \, t^{\frac{z}{2}-1} \,
\Big[ S(t) - 1 \Big] \, ,
\end{equation}
we can express the coefficients $a_p$ in terms of the Riemann $\zeta$-function as
\begin{equation}
a_p = \frac{c_p}{8 \pi^{2p+\frac{3}{2}}} \, \Gamma( 2p-\mbox{$\frac{1}{2}$}) \zeta(4p -1) \, .
\end{equation}
The final representation for the quark condensate in the chiral limit and $|qH| \ll T^2$ thus reads 
\begin{eqnarray}
\label{finqcondAnalytic}
\frac{\langle {\bar q} q \rangle}{\langle 0 | {\bar q} q | 0 \rangle} & = & 1 - \frac{1}{8 F^2} \, T^2 + \frac{|qH| \log 2}{16 \pi^2 F^2}
\nonumber \\
& & + \Bigg\{ \frac{|I_{\frac{1}{2}}|}{8 \pi^{3/2} F^2} \sqrt{\epsilon}
- \frac{\log 2}{16 \pi^2 F^2} \, \epsilon
+\frac{\zeta(3)}{384 \pi^4 F^2} \, \epsilon^2
- \frac{7 \zeta(7)}{98 304 \pi^8 F^2} \, \epsilon^4 + {\cal O}(\epsilon^6) \Bigg\} \, T^2 \nonumber \\
& & + {\cal O}(T^4) \, .
\end{eqnarray}

\subsection{Straightforward Derivation of the Leading Terms}

In order to readily derive the leading terms in the quark condensate in the chiral limit and $|qH| \ll T^2$, we consider the relevant Bose
function
\begin{equation}
{\tilde g_1}(0, T, H) = \frac{|qH|}{16 \pi^2} \, {\int}_{\!\!\! 0}^{\infty} dt \, t^{-1} \Big( \frac{1}{\sinh(t)} - \frac{1}{t} \Big)
\Bigg[ S\Big( \frac{|qH|}{4 \pi T^2 t} \Big) -1 \Bigg] \, ,
\end{equation}
that we write as
\begin{eqnarray}
\label{decompIdentity}
{\tilde g}_1(0, T, H) & = & - \frac{|qH|}{16 \pi^2} \, {\int}_{\!\!\! 0}^{\infty} dt \, t^{-1} \Big( \frac{1}{\sinh(t)} - \frac{1}{t} \Big)
\nonumber \\
& & + \frac{\sqrt{|qH|}T}{8 \pi^{3/2}} \, {\int}_{\!\!\! 0}^{\infty} dt \, t^{-1/2} \Big( \frac{1}{\sinh(t)} - \frac{1}{t} \Big)
S\Big( \frac{4 \pi T^2}{|qH|} t \Big) \, .
\end{eqnarray}
Note that the Jacobi theta function
\begin{equation}
S(z) = \sum_{n=-\infty}^{\infty} \exp(- \pi n^2 z)
\end{equation}
satisfies the identity 
\begin{equation}
S(z) = \frac{1}{\sqrt{z}} \, S(\frac{1}{z}) \, .
\end{equation}
The integral in the first line, Eq.~(\ref{decompIdentity}), is known analytically,
\begin{equation}
{\int}_{\!\!\! 0}^{\infty} \, dt t^{-1} \Big( \frac{1}{\sinh(t)} - \frac{1}{t} \Big) = - \log 2 \, ,
\end{equation}
and gives rise to the correction linear in $\epsilon$ in Eq.~(\ref{finqcond}). Regarding the second line, Eq.~(\ref{decompIdentity}), in
the limit $|qH| \ll T^2$, all contributions in the Jacobi theta function -- except $n$=0 -- are exponentially suppressed: the
corresponding integral hence reduces to $I_{\frac{1}{2}}$, Eq.~(\ref{I12}), and we immediately obtain the leading temperature-dependent
term in the weak magnetic field expansion $|qH| \ll T^2$ in the chiral limit,
\begin{equation}
\frac{\sqrt{|qH|} T}{8 \pi^{3/2} F^2} \, |I_{\frac{1}{2}}| \, .
\end{equation}

\end{appendix}

\end{document}